# Hand and Arm Gesture-based Human-Robot Interaction: A Review

XIHAO WANG

Technical University of Munich, School of Computation, Information and Technology

HAO SHEN

Fortiss GmbH, Machine Learning Laboratory

HUI YU

Fujian Science & Technology Innovation Laboratory for Optoelectronic Information of China

JIELONG GUO

Fujian Science & Technology Innovation Laboratory for Optoelectronic Information of China

XIAN WEI

Fujian Science & Technology Innovation Laboratory for Optoelectronic Information of China

The study of Human-Robot Interaction (HRI) aims to create close and friendly communication between humans and robots. In the human-center HRI, an essential aspect of implementing a successful and effective HRI is building a natural and intuitive interaction, including verbal and nonverbal. As a prevalent nonverbally communication approach, hand and arm gesture communication happen ubiquitously in our daily life. A considerable amount of work on gesture-based HRI is scattered in various research domains. However, a systematic understanding of the works on gesture-based HRI is still lacking. This paper intends to provide a comprehensive review of gesture-based HRI and focus on the advanced finding in this area. Following the stimulus-organism-response framework, this review consists of: (i) Generation of human gesture(stimulus). (ii) Robot recognition of human gesture(organism). (iii) Robot reaction to human gesture(response). Besides, this review summarizes the research status of each element in the framework and analyze the advantages and disadvantages of related works. Toward the last part, this paper discusses the current research challenges on gesture-based HRI and provide possible future directions.

## 1 INTRODUCTION

With the development of robot systems, robots have already stepped into our daily life. They support human users in various areas, such as industry [26, 13, 14], retail [9, 59], home service [10, 19], entertainment [39], and caretaker [4, 52, 22, 15]. The importance of constructing a natural and intuitive interaction between humans and robots is self-evident. Inspired by the human-human interaction (HHI), the communication approach in human-robot interaction (HRI) is also through verbal and nonverbal channels [2]. Within these interaction channels, nonverbal communication is an unspoken dialogue that creates shared meaning in social interactions [3]. As one important aspect of the HRI, nonverbal interaction can help the robot provide communicative functionality that is natural and intuitive to their human interaction partners [44].

Gesture, as a popular nonverbal HRI, includes hand and arm movement, body behavior, facial emotional expression, and gaze shifts [37]. Hand and arm gestures are generally defined as the upper body limb's significant movement and produce an expression of feeling or rhetoric [46]. Among the HRI approaches, hand and arm gesture-based interaction is one of the most important nonverbal interaction methods. It is due to the irreplaceable advantages brought by gesture-based in HRI for several reasons: (i) Gesture-based interaction is one of the critical elements of human communication, and it offers a natural and intuitive approach to HRI spontaneously [42]. (ii) Gestures could help human users to convey the specific intent, which offers a more efficient and convenient communication than verbal in specific scenarios. For example, it is easy to use a pointing gesture to indicate the target object to the robot in a pick-and- place scenario, whereas hard to do so verbally. (iii) Gesture-based interaction plays an essential role in an exceptional environment where verbal communication is limited. For instance, gesture-based interaction has become the most efficient method for divers to interact with their underwater robot [16]. A similar also appears in other verbal HRI limited situations, including outer space, noisy factories, and quiet libraries. (iv) For the elder and children who are inexperienced users, hand and arm gesture-based HRI allows them to communicate with the robots not requiring any knowledge of the robots' internal states [43]. Despite the above explicit advantages, hand and arm gesture-based HRI also implicitly supports a friendly and safe human-robot relationship. One of the parts is that robots could predict the motion of humans depending on implicit gesture-based communication, which significantly improves teamwork efficiency and safety in human-robot collaboration places [2]. On the other hand, the communication based on hand and arm gesture increases the robot's anthropomorphic degree, which provokes a larger possibility of future interaction intentions from the human [42]. Therefore, gesture-based HRI not only offers an intuitive and natural communication method but also provides an approach to enhance the interaction between humans and robots.

However, knowledge about gesture-based HRI is still fragmented in several research directions. Even though some overview studied on gesture recognition during HRI [24, 5, 45], they drew their sight on the technology used to recognize the human gesture. Those papers introduced human gesture recognition through multiple methodologies, including the model, network, and framework in computer vision. However, gesture-based HRI is a complicated task, while gesture recognition is only one of the parts that the robot percepts the intent of the human. Hand and arm gesture-based HRI should be considered as a complete process, including human operation and robots' reactions. In existing research, the reaction of robots is usually not considered during the interaction, and the effect of gesture-based HRI is also studied individually. Thus, it is necessary to summarize the research and do a literature review on a novel and comprehensive sight.

In this paper, we propose the framework based on the stimulus-organism-response (SOR) paradigm [18] to establish a comprehensive introduction of hand and arm gesture-based HRI. Within this framework, human gesture generation is the stimulus (S) that is captured through gesture recognition by the robot as a mediator (O), and the robot reacts to this gesture as a response (R) [53].

Following the framework described in Figure 1, we go through the process of the gesture-based HRI from the semantics of human gesture (Sec. 2), the gesture recognition from the robot (Sec. 3), and the reaction of the robot during the interaction (Sec. 4). Last but not least, we discuss some of the shortcomings of existed works and investigate some potential research that could be carried out in the future (Sec. 5). Our main contributions are: (i) We provide a novel sight to categorize hand and arm gesture-based HRI into three phases based on the stimulus-organism-response paradigm: human gesture generation, gesture perception, and robot reaction. Unlike specifically focusing on one part of the HRI, we offer a comprehensive understanding of the complete process of the hand and arm gesture-based HRI. (ii) We objectively summarize the research status of each element in our framework and analyze the advantages and disadvantages of mentioned research works. (iii) By summarizing the current research, we raise various current challenges in hand and arm gesture-based HRI and discuss the potential future directions.

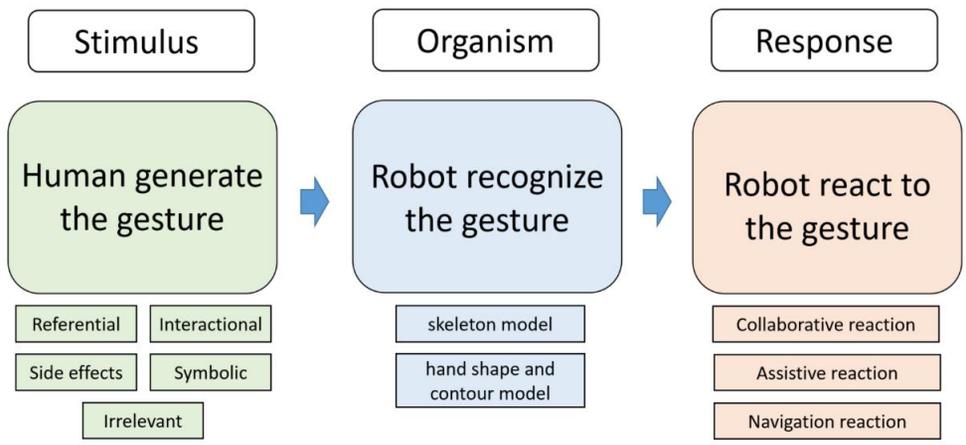

Figure 1: Conceptual framework for categorizing works on gesture-based human-robot interaction.

The paper is organized as follows. In Sec. 2, we present the study of the semantics of human gestures during hand and arm gesture-based HRI. Then, in Sec. 3, we introduce the gesture recognition model of the robot. In Sec. 4, we illustrate the reaction from the robot after the robot received the comment from the human. Finally, in Sec. 5, we propose several suggestions for the potential research direction on gesture-based HRI.

## 2 SEMANTIC OF HUMAN GESTURE

Before considering robot recognition and reaction, we need to discuss how humans make gestures. According to our framework, the stimulus is the gesture generated by humans. As mentioned above, hand and arm gestures are movement expressing feeling and rhetoric. Thus, as an expression and rhetoric, categorizing the gesture in gesture-based HRI and understanding their intention becomes a pre-request work.

The gesture generated by human users expresses their intention and meaning to stimulate the robot and start the process of HRI. The categorization of the human gesture could clearly define the stimulation factors, which improves the robot's understanding of human users' wishes. According to the concept of mindless transfer in the Computer-As-Social-Actor paradigm [28], it indicates the fact that humans would mindlessly employ HHI behaviors when they interact with the robot. Besides, compared with the HRI, human-computer interaction (HCI) has been studied longer. Thus, inspired by HHI [27] and HCI [21], hand and arm gesture in HRI could be categorized into five classes [29], and we assigned them with related works as following: (i) Referential/pointing gesture: These are used to refer to or to indicate objects or location. The related scenarios are about referring to an object for the robot to pick [23, 38, 41, 51], and referring a place for the robot to arrive[4, 36] (ii) Interactional gesture: These are used to regulate interaction with a partner, i.e., to initiate, maintain, invite, synchronize, organize, or terminate a particular interactive/cooperative behavior. The related tasks are about conducting handover task [26, 50], holding task [30], and assembly task[13, 14, 47]. (iii) Side effects of expressive behavior: When humans communicate with others, hands and arms' motions occur as part of the overall communicative behavior. Nevertheless, those gestures do not have specific interactive, communicative, symbolic, or referential roles. (iv) Symbolic gesture: gestures are associated with firm meaning in our cultural environment and are frequently used in everyday experiences. The related scenarios are about directing the movement of robot [10, 12, 16, 58], presenting human's opinion [20, 26, 38], and presenting specific sign language [16, 17, 32]. (v) Irrelevant/manipulative gesture: gestures which are neither communicative nor socially interactive. They are rather instances of effects of human motion. Cases include, i.e., the motion of the arms and hands when walking, tapping of the fingers, playing with a paper clip, etc. The detail about the categorization of three kinds of interaction is presented in the Table 1

Table 1: The gestures categorization in different interaction

| Interaction | Categorization | Explanation |
|---|---|---|
| Human-Human Interaction [27] | Iconic | Representing the image of concrete actions |
| | Beats | Performing hand movement rhythmically |
| | Deictic | Pointing target and object |
| | Metaphoric | Representing abstract idea and information |
| Human-Computer Interaction [21] | Deictic | Pointing to establish the identity or spatial location |
| | Manipulative | Represents the intention to control some entity |
| | Semaphoric | Represents a sign or signal to communicate information |
| | Gesticulation | Gesture combines with conversational speech interfaces |
| | Language | Linguistically based and performs a series of individual signs |
| | Multiple | Combining more than one kind of gesture style mentioned |
| Human-Robot Interaction [29] | Referential | Referring to or to indicate objects or location |
| | Interactional | Regulate interaction with a partner |
| | Side effects | Motions occur as part of the overall communication |
| | Symbolic | Associated with firm meaning in cultural environment |
| | Irrelevant | They are rather instances of effects of human motion |

**Major finding**: Although the research of HCI [21] and HHI [20] could not be the same as HRI [29], humans also would like to generate their gestures by their most familiar style. Thus, we can still discover that they contain a certain internal similarity. In the rest of the paper, we would name the gestures according to the HRI classification [29]. Regarding the side effects of expressive behavior and Irrelevant/manipulative gestures, even though they play a blurred role in HRI tasks, they are also essential to predict human intention and emotion. Efficient categorization can improve the legibility of human gestures, so that the robot would be better to understand and predict human intention.

## 3 GESTURE RECOGNITION OF THE ROBOT

In the SOR model, an organism refers to any "internal processes and structures intervening between stimuli external to the person and the final actions, reactions or responses emitted" [18]. For this review, the organism is represented by robots' recognition of human gestures. Depending on the recognized objection, we categorize the methods into two classes and illustrates their advantages:

*Skeleton model*: Regarding a significant number of the gesture-based HRI, a detailed model of the human's hand may not be useful. The Skeleton model approach uses humans' skeleton to identify the pose of the human body [24]. The skeleton model could simplify the human body's posture information and present the meaning of the gesture. For instance, the head position [40, 54] or the relative position of the torso [26] could become a promised reference to recognize the gesture. Besides, the skeleton model is less influenced by the surrounding environment and the human body's movement. Thus, the skeleton model is usually used in navigation and collaboration tasks depending on the above benefit [11, 57]. The skeleton model is widely employed in gesture-based HRI research with the advantage of simplicity and reliability.

*Hand shape and contour model*: Unlike the skeleton model, the hand shape and contour method is adopted to extract the detail of the gesture. The detailed information includes the fingertip, the palm, and the number of fingers [16, 17]. Due to this method only focusing on the hand part, it usually employs the skin color [31, 48] and depth information [55, 56] to locate the place of the hand. Depending on the advantage of extracting enriched gesture semantic information, the hand shape and contour approach is also frequently utilized in gesture-based HRI.

**Major finding**: The research on gesture recognition provided many approaches to identify the gesture involved in HRI. Because of the visual sensor's outstanding property, most of the works capture gestures via image data. Besides, contact-based sensing, such as Inertial Measurement Unite [7], also attracts many researchers' interest due to its mobility and economic cost. With the development of the deep learning algorithm, efficient computation and enriched datasets encourage the researchers to employ the end-to-end prediction model to implement gesture recognition rather than one explicit model. In the works of [48, 49], the authors used static hand gestures from the Innsbruck Multi-view Hand Gestures database [49] to instruct the robot. Depending on the benefit of efficiency and comprehension, deep learning-based methodology [34, 35] gradually becomes the mainstream of gesture recognition during hand and arm gesture-based HRI.

## 4  REACTION FROM THE ROBOT

In the SOR paradigm, response is a person's reaction to a stimulus [18]. Regarding gesture-based HRI, the robot reaction depends on human users' gestures. After the human users' gestures or intentions are recognized, the robot needs to understand, interact, and share information with people in real-world human-centered environments [33]. In this section, we categorize the robot reaction into three different scenarios described in Table 2 and illustrate the reaction mechanism.

### 4.1  COLLABORATIVE REACTION

Human-robot collaboration is an essential research domain of HRI, where the robot and human work together to complete a common task [24]. To overcome the restricted environment, hand and arm gestures become a suitable HRI communication approach. The collaboration reaction could be divided into three main scenarios: industry scenario, daily life scenario, and exploration scenario.

In the industry aspect, the robot plays the role of the teammate of the human workers. The works [13, 14, 47] studied the task that the human worker to collaborate with the single-arm manipulators, such as the Barrett WAM arm. The robot is designed to respond to the ten kinds of gestures that usually occur in industry assembly scenarios [13], and each kind of gesture corresponds to a specific reaction. For instance, the twist gesture is associated with the rotating part's intention. Moreover, Mazhar et al. [26] developed a framework for handover tasks by the BAZAR robot. According to different collaboration requirements, the reaction mechanism could be a hard-coded or dynamic model.

Table 2: The brief summary of reaction categorization and key related works

| Reaction Scenario | Categorization | Reference |
|---|---|---|
| Collaborative Reaction | Industry collaboration | [13], [14], [26], [47] |
| | Daily life collaboration | [38], [50], [51] |
| | Extremely environment exploration | [16], [25] |
| Assistive Reaction | Elder care | [4] |
| | Laboratory assistance | [8] |
| | Medical treatment | [17], [60] |
| | Children education | [6], [39] |
| Navigation Reaction | Transportation | [10], [58] |

Regarding the daily life scenario, Quintero et al. [38] explore the collaboration work with humans and the 7DOF WAM arm using gestures to select, pick, and drop objects at different locations. In their experiment, the human operator conducts the semaphore's gesture to interact with the robot for making a pizza. The group of Shukla et al. [50, 51] proposed the Proactive Incremental Learning framework, which supervised the robot to study the association between human hand gestures and the intended robot manipulation actions. Due to the flexible requirement of daily life, the reaction mechanism is usually the dynamic model.

Under the exploration scenario, the robot is a reliable teammate when the human is doing the exploration task. With the robot's collaboration, the dangerous task in exploration would not need the human to implement personally. In the research from Islam et al. [16], they presented a real-time gesture-based interaction method for autonomous underwater robots in human-robot collaborative tasks. Depending on the two-hand combination gesture, the robot implements the actions, including hovering, diving down, and floating up. Likely, the research from Liu et al. [25] proposed a gesture-based HRI system for astronauts with extravehicular activities. They designed to use a glove sensor inside the spacesuits to interface with the assistant robot. Because the robot in the exploration scenario needs to real-time react with the human operator, the reaction mechanism is usually the dynamics model.

## 4.2 ASSISTIVE REACTION

With the population growing, elderly care has gradually become an urgent problem in most developed countries. The assistive reaction involved in the robot needs to aid people, such as the aged, patient, or even someone very busy. Because those people cannot interact like normal adults, the gesture could help them convey their intention to the robot. Canal et al. [4] proposed a gesture-based HRI system to assist the aged. This system is composed of NAO and Wifibot robots. The system would recognize the pointing and wave hand gestures by the Kinect sensor and classify them by a trained deep neural network. Depending on the skeleton angle, the system could detect the desired place from pointing gestures and uses Wifibot to send the NAO robot there. Consequently, the NAO robot would operate the pick-and-place task toward the target object.

Regarding assistance for the people who have a special job, Gestonurse [17] is a multimodal robotic scrub nurse for assisting the doctor who needs to focus on surgery. According to the finger gesture, the Gestonurse could pick related surgery equipment, such as the scalpel, scissors, etc.

About the research of [8], the authors conducted a gesture-based HRI laboratory scenario. They set up with DLR LWR three-arms manipulator to deal with dangerous liquid.

Moreover, the assistive reaction could also have a unique effect on medicine, education, and entertainment. In terms of medical treatment, the robot uses gestures to imitate the children's behavior to treat children with autism [60]. The research showed that the robot gesture imitation system draws more attention from children with autism and presented markable effectiveness. In [39], the author introduces several games, like rock-paper-scissors. [6] presented a study about using the robot's gesture to teach children the second language vocabulary. In this research, they designed the NAO robot to perform the iconic gesture related to the vocabulary's hint. The experiment result indicated that robotic gesture positively affects the long-term memorization of novel words. And it also showed that children have a higher level of engagement during learning activities.

### 4.3 NAVIGATION REACTION

Navigation is one of the tasks that is difficult to achieved through verbal communication. Thus, gestures are usually used to navigate the robot to move to a specific place. Human users could interact with the mobile robot to move where they want to go without any expert knowledge. Navigation gestures are natural and intuitive, such as turn left, turn right, go straight, and stop.

In the research of [10, 58], the researchers proposed the gesture-based navigation system for HRI. The mobile robot recognizes the angle of the human user's hand and arm skeleton angle by the Kinect sensor [10]. The system used a trained Hidden Markov Model to classify the recognized gesture into predefined gestures, thus human users could conduct the mobile robot to carry out the following human service in the office environment. Moreover, depending on the transport gesture, the robot could implement a delivery service without the human user's navigation. Likely, in [58], the authors used the Inertial Measurement Unite sensor to capture the human user's hand and arm gestures.

**Major finding**: The robot's reaction to gesture-based HRI presents the versatile ability in various scenarios. These studies affirm that the robot's reaction could be initiative or passive. In the example of assistive reaction, the robot has to wait for the human user to give commands. In terms of collaboration and navigation reaction, the robot can initially react depending on the human's gesture and posture. Unlike the human-center interaction, in the situation of social behavior reaction, the robot which imitates human behavior shows the potential to operate the robot-center gesture-based interaction with a human. As the aspect of the reaction mechanism, the learning-based reaction presents outstanding performance during gesture-based HRI. The robot's learning and prediction ability provide a better service during the HRI process. However, the research mentioned above still builds on the desired scenarios rather than the unfamiliar environment.

### 5 DISCUSSION

In the sections above, we discussed the process of human-center gesture-based HRI. Although there is a large amount of research investigated in each domain, there are still some gaps in the gesture-based HRI. We propose several suggestions for further research on gesture-based HRI.

*Suggestion 1*: Gesture-based interaction is a kind of communication that is easily influenced by the interaction scenario and context. In the present research [13], [14], [26], [47], the robot collaborates with only one human partner. There has been little analysis or discussion about the scenario of multiple workers' collaboration. Thus, increasing the environment's robustness is one of the potential directions of hand and arm gesture-based HRI.

Suggestion 2: Gesture interaction is natural and intuitive between humans and robots. But verbal interaction also provides irreplaceable function. In the research of [44], the authors suggest that the robot is evaluated more positively when nonverbal behaviors such as hand and arm gestures are displayed along with speech. Thus, co-verbal gesture-based HRI may provide better communication than only nonverbal interaction.

*Suggestion 3*: Although gesture-based interaction is an intuitive and natural communication approach, the research [1] also indicates the incorrect responses of gestures are more than gestures with speech. It is necessary to raise a correction mechanism through hand and arm gestures during the HRI. For instance, the robot would conduct an inquiry if the pick object is right or not depending on the human's thumb up or down gesture [38]. Due to the gesture's larger mental workload, there has considerable research needs to be done to provide an effective correction during the interaction.

## 6 CONCLUSION

In this paper, we conduct a review of hand and arm gesture-based HRI. Hand and arm gesture interaction plays an important role in human-robot communication. Because of its intuitive and natural property, all people could accept gesture-based interaction with the robot, even those with little knowledge of robotics. However, far too little attention has been paid to aggregate research works, which fragmented into several domains. Based on the proposed SOR paradigm, we provide a comprehensive framework to categorize current research into human gesture generation, gesture recognition, and robot behavior reaction. Depending on the status of the domain, we raise various current challenges, such as noise robustness, co-verbal gesture-based HRI, and effective correction mechanism. We believe that our review would stimulate the research on hand and arm gesture-based HRI to achieve further progress in the future.